\documentstyle[12pt,epsf]{article}
\expandafter\ifx\csname amssym12.def\endcsname\relax \else\endinput\fi
\expandafter\edef\csname amssym12.def\endcsname{%
       \catcode`\noexpand\@=\the\catcode`\@\space}
\catcode`\@=11
\def\undefine#1{\let#1\undefined}
\def\newsymbol#1#2#3#4#5{\let\next@\relax
 \ifnum#2=\@ne\let\next@\msafam@\else
 \ifnum#2=\tw@\let\next@\msbfam@\fi\fi
 \mathchardef#1="#3\next@#4#5}
\def\mathhexbox@#1#2#3{\relax
 \ifmmode\mathpalette{}{\m@th\mathchar"#1#2#3}%
 \else\leavevmode\hbox{$\m@th\mathchar"#1#2#3$}\fi}
\def\hexnumber@#1{\ifcase#1 0\or 1\or 2\or 3\or 4\or 5\or 6\or 7\or 8\or
 9\or A\or B\or C\or D\or E\or F\fi}
\font\tenmsa=msam10 scaled\magstep1
\font\sevenmsa=msam7 scaled\magstep1
\font\fivemsa=msam5 scaled\magstep1
\newfam\msafam
\textfont\msafam=\tenmsa
\scriptfont\msafam=\sevenmsa
\scriptscriptfont\msafam=\fivemsa
\edef\msafam@{\hexnumber@\msafam}
\mathchardef\dabar@"0\msafam@39
\def\dashrightarrow{\mathrel{\dabar@\dabar@\mathchar"0\msafam@4B}}
\def\dashleftarrow{\mathrel{\mathchar"0\msafam@4C\dabar@\dabar@}}

\def\ulcorner{\delimiter"4\msafam@70\msafam@70 }
\def\urcorner{\delimiter"5\msafam@71\msafam@71 }
\def\llcorner{\delimiter"4\msafam@78\msafam@78 }
\def\lrcorner{\delimiter"5\msafam@79\msafam@79 }
\def\yen{{\mathhexbox@\msafam@55 }}
\def\checkmark{{\mathhexbox@\msafam@58 }}
\def\circledR{{\mathhexbox@\msafam@72 }}
\def\maltese{{\mathhexbox@\msafam@7A }}
\font\tenmsb=msbm10 scaled\magstep1
\font\sevenmsb=msbm7 scaled\magstep1
\font\fivemsb=msbm5 scaled\magstep1
\newfam\msbfam
\textfont\msbfam=\tenmsb
\scriptfont\msbfam=\sevenmsb
\scriptscriptfont\msbfam=\fivemsb
\edef\msbfam@{\hexnumber@\msbfam}

\def\widehat#1{\setbox\z@\hbox{$\m@th#1$}%
 \ifdim\wd\z@>\tw@ em\mathaccent"0\msbfam@5B{#1}%
 \else\mathaccent"0362{#1}\fi}
\def\widetilde#1{\setbox\z@\hbox{$\m@th#1$}%
 \ifdim\wd\z@>\tw@ em\mathaccent"0\msbfam@5D{#1}%
 \else\mathaccent"0365{#1}\fi}
\font\teneufm=eufm10 scaled\magstep1
\font\seveneufm=eufm7 scaled\magstep1
\font\fiveeufm=eufm5 scaled\magstep1
\newfam\eufmfam
\textfont\eufmfam=\teneufm
\scriptfont\eufmfam=\seveneufm
\scriptscriptfont\eufmfam=\fiveeufm

\csname amssym.def\endcsname
\newif{\ifcomentarios}
\comentariosfalse
\renewcommand{\theequation}{\thesection.\arabic{equation}}

\newcommand{\zerarcounters}
{
\setcounter{equation}{0}
\setcounter{theorem}{0}
}

\newcommand{\be}{\begin{equation}}
\newcommand{\ee}{\end{equation}}
\newcommand{\bma}{\begin{displaymath}}
\newcommand{\ema}{\end{displaymath}}
\newcommand{\bc}{\begin{center}}
\newcommand{\ec}{\end{center}}
\newcommand{\text}{\rm}

\newcommand{\uflex}
{{\scriptstyle {\raise 9pt\hbox{$\backslash$}\,\!\!\!\!\!\Bigg\vert}}}
\newcommand{\ncm}{\newcommand}

\ncm{\rncm}{\renewcommand}
\ncm{\id}{{\bf 1}}
\ncm{\beq}{\begin{equation}}
\ncm{\eeq}{\end{equation}}
\ncm{\ba}{\begin{array}}
\ncm{\bea}{\begin{eqnarray}}
\ncm{\beanon}{\begin{eqnarray*}}
\ncm{\ea}{\end{array}}
\ncm{\eea}{\end{eqnarray}}
\ncm{\eeanon}{\end{eqnarray*}}
\ncm{\fns}{\footnotesize}
\ncm{\setc}[1]{\setcounter{equation}{#1}}
\newcounter{eqnr}
\newenvironment{eqnarrayabc}{\stepcounter{equation}
  \setcounter{eqnr}{\value{equation}}\setc{0}
  \rncm{\theequation}{\thesection.\arabic{eqnr}\alph{equation}}
  \begin{eqnarray}}{\end{eqnarray}\setc{\value{eqnr}}}
\ncm{\eqboxabc}[3]{\newline\parbox[t]{1.5cm}{#1}\hfill
  \parbox[b]{12cm}{\begin{eqnarray*} #3\end{eqnarray*}}\hfill
   \parbox[b]{1.5cm}{\vspace{-0.0cm}\begin{eqnarrayabc}#2\end{eqnarrayabc}}\newline}
\ncm{\eqbox}[2]{\newline\parbox{1.5cm}{#1}\hfill
  \parbox{12cm}{\beanon #2\eeanon}\hfill
  \parbox{1cm}{\bea\eea}\newline}
\ncm{\nr}[1]{\parbox{1cm}{\begin{eqnarrayabc}#1\end{eqnarrayabc}}\\}

\ncm{\kal}[1]{\mbox{$\cal #1 $}}
\ncm{\mrk}[1]{\!\!\! #1 \!\!\!} 
\ncm{\qed}{\hspace*{0.4cm}\rule{0.24cm}{0.24cm}}  
\ncm{\mbold}[1]{\mbox{\boldmath $ #1 $}}   
\ncm{\bm}{\mbold}
\ncm{\str}{\stackrel}
\ncm{\sub}{\subset}
\ncm{\e}{\varepsilon}
\ncm{\ka}{\kappa}
\ncm{\inputc}[1]{\begin{center}\input{#1}\end{center}}
\ncm{\lto}{\longrightarrow}
\ncm{\x}{\times}
\ncm{\bmm}{\bm{\cal M}}
\ncm{\cp}{{\bf P}}    
\ncm{\bfp}{{\bf P}}
\ncm{\bmi}{\bm{i}}
\ncm{\bmom}{\bm{\om}}
\ncm{\bmOm}{\bm{\Om}}
\ncm{\res}{\restriction}
\ncm{\bmL}{\bm{\cal L}}
\ncm{\bmell}{\bm{\ell}}
\ncm{\bmE}{\bm{\cal E}}
\ncm{\bme}{\bm{e}}
\ncm{\bmpi}{\bm{\pi}}
\ncm{\bmr}{\bm{r}}
\ncm{\bmsigma}{\bm{\sigma}}
\ncm{\wt}{\widetilde}
\newcommand{\beaa}{\begin{eqnarray}}
\newcommand{\eeaa}{\end{eqnarray}}

\begin{document}
\input{epsf.tex}

\author{{\bf Oscar Bolina}\thanks{Supported by FAPESP under grant
97/14430-2. {\bf E-mail:} bolina@math.ucdavis.edu} \\
%EndAName
Department of Mathematics\\
University of California, Davis\\
Davis, CA 95616-8633 USA\\
\and {\bf J. A. da Silva Neto}\thanks{Supported by FAPESP under 
grant 97/01003-9. {\bf E-mail:} jantonio@fma.if.usp.br} \\ 
%EndAName
Instituto de F\'{\i}sica \\
Universidade de S\~ao Paulo \\
Caixa Postal 66318 \\
05315-970 S\~ao Paulo, Brasil \\
}
\title{\vspace{-1in}
{\bf Finite Rotations}}
\date{}
\maketitle
\begin{abstract}
\noindent
We present an elementary discussion of two basic properties of 
angular displacements, namely, the anticommutation of finite 
rotations, and the commutation of infinitesimal rotations, and 
show how commutation is achieved as the angular displacements get
smaller and smaller.

\noindent
{\bf Key words:} Finite Rotations, Commutivity \hfill \break
%\vfill
\end{abstract}

%\newpage

%%%%%%%%%%%%%%%%%%%%%%%%%%%%%%%%%%%%%%%%%%%%%%%%%%%%%%%%%%%%%%%%%%%%%%%%%%%%%%
%1
%%%%%%%%%%%%%%%%%%%%%%%%%%%%%%%%%%%%%%%%%%%%%%%%%%%%%%%%%%%%%%%%%%%%%%%%%%%%%%

\section{Introduction}
\zerarcounters
Even though finite rotations can be represented by a magnitude (equal
to the angle of rotation) and a direction (that of the axis of rotation),
they do not act like vectors. In particular, finite rotations do not
commute: The summation of a number of finite rotations, not about the
same axis, is dependent on the order of addition.
\newline 
The anticommutivity property of finite rotations is made clear 
in introductory texts by showing that two successive finite 
rotations, when effected in different order, produce different 
final results \cite{Slater}. 
\newline
However, when rotations are small -- indeed, infinitesimal --,
the combination of two or more individual rotations is unique,
regardless of the order they are brought about (This fact allows 
for the definition of angular velocity as the time-derivative of 
an angular coordinate [2, p.675]).
\newline
Here we show how the order rotations are carried out becomes
irrelevant -- that is, rotations become commutative -- as the 
angles of rotation diminish.
\newline 
In Fig. 1 we have represented two successive rotations of a rigid 
body. The first rotation is around the axis {\it OZ} through an
angle $\phi$, which takes {\it OA} into {\it OB}. For simplicity,
we take the plane {\it OAB} to be the horizontal {\it XY} plane. The
second rotation is around the axis {\it OX} through an angle $\theta$,
which takes {\it OB} into {\it OC}. 
\newline
Since the angle between the axes {\it OB} and the axis of rotation
{\it OX} is not $90^{\circ}$, the plane {\it OBC} cuts the plane
{\it XY} at an angle. Let this angle be $\beta$, represented in
Fig. 1 as the angle formed by the sides {\it PQ} and {\it QR} of 
the triangle {\it PQR}. 
\newline
After these two rotations, the initial point {\it A} is brought to 
the final position {\it C}. This same final result can be accomplished
by just one rotation through an angle $(A,C)$ around an axis
perpendicular to both {\it OB} and {\it OC}. 
\newline
To obtain a relation between the angles $\phi, \theta, \beta$ and 
{\it (A,C)}, we have drawn in Fig. 2 four triangles, derived from
Fig. 1, which are relevant to our analysis.
\newline
From the two right triangles {\it OQP} and {\it OQR}, we have the
relations 
\beq
\cos \phi=\frac{OQ}{OP}, \;\;\;\;\;\;\;\;\;\;
\sin \phi=\frac{PQ}{OP} \label{1}
\eeq
and
\beq
\cos \theta=\frac{OQ}{OR}, \;\;\;\;\;\;\;\;\;\;
\sin \theta=\frac{QR}{OR}. \label{2}
\eeq
The law of cosines applied to the triangles {\it OPR} and {\it PQR}
yields
\beq
{PR}^{2}={OP}^{2}+{OR}^{2} -2 (OP)(OQ)\cos(A,C)
\label{3}
\eeq
and
\beq
{PR}^{2}={PQ}^{2}+{QR}^{2} -2 (PQ)(QR)\cos\beta.
\label{4}
\eeq 
Substituting for {\it PQ} and {\it QR} their values given in
(\ref{1},\ref{2}), Eq. (\ref{4}) becomes
\beq
{PR}^{2}={OP}^{2}\sin^{2}{\phi} + {OR}^{2}\sin^{2}{\theta} 
-2 (OP)(OR)\sin{\phi} \sin{\theta} \cos\beta \label{5}
\eeq 
On equating expressions (\ref{3}) and (\ref{5}) for {\it PR},
using (\ref{1},\ref{2}), we get
\beq
\cos(A,C)=\cos{\phi}\cos{\theta} + \sin{\phi}\sin{\theta} \cos\beta
\label{6}
\eeq
Now we effect the rotations in the reverse order, taking the first
rotation around the axis {\it OX} through an angle $\theta$,
followed by a rotation around the axis {\it OZ} through an angle
$\phi$. In this case, the point {\it A} moves to the new final
position {\it E}, instead of {\it C}, as indicated in the sketch 
accompanying Fig. 1. The same final result can again be accomplished
by just one rotation, now through an angle $(A,E)$ around an axis
perpendicular to both {\it OD} and {\it OE}. 
\newline
A moment's reflection shows that the relation between the
angles now is analogous to (\ref{6}), with no need to repeat
the above procedure. The cosine of the angle $(A,E)$ is given by  
$\cos(A,E)=\cos{\theta}\cos{\phi} + \sin{\theta}\sin{\phi} 
\cos{\beta'}$, with the difference that $\beta'$ is the angle the
plane {\it AOD} makes with the horizontal plane {\it DOE}. Thus
we have 
\beq
\cos(A,C)-\cos(A,E)=\sin{\phi}\sin{\theta}(\cos{\beta}-\cos{\beta'})
\label{8}
\eeq
If we set $\beta'=\beta+\Delta \beta$, and expand $\cos\beta'$ 
in (\ref{8}) we get
\beq
\cos(A,C)-\cos(A,E)=\sin{\phi}\sin{\theta}[\cos{\beta}
(1-\cos{\Delta \beta}) +\sin{\beta} \sin{\Delta \beta}]
\label{9}
\eeq
To see how commutivity is obtained when the angles involved are small,
we use that, for $x \ll 1$, $\sin{x} \approx x$, $\cos{x} \approx 1$.
Eq. (\ref{9}) becomes
\beq
\cos(A,C)-\cos(A,E) \approx \phi~ \theta~ {\Delta \beta}~ \sin{\beta} 
\label{10}
\eeq
This means that the difference between the two final positions 
vanishes more rapidly than either of the single rotations.
\newline
{\it Remark:} It is not necessary to assume that $\Delta \beta$ is 
small. Our result holds whether it is small or not, since in
(\ref{8}) we could have simplified our analysis by using that $\mid
\cos{\beta}-\cos{\beta'} \mid \leq 2$, and getting the same conclusion
above.

%%%%%%%%%%%%%%%%%%%%%%%%%%%%%%%%%%%%%%%%%%%%%%%%%%%%%%%%%%%%%%%%%%%%%%%%%%%%%
%BIBL
%%%%%%%%%%%%%%%%%%%%%%%%%%%%%%%%%%%%%%%%%%%%%%%%%%%%%%%%%%%%%%%%%%%%%%%%%%%%%

\newpage
\begin{figure}
\centerline{
\epsfbox{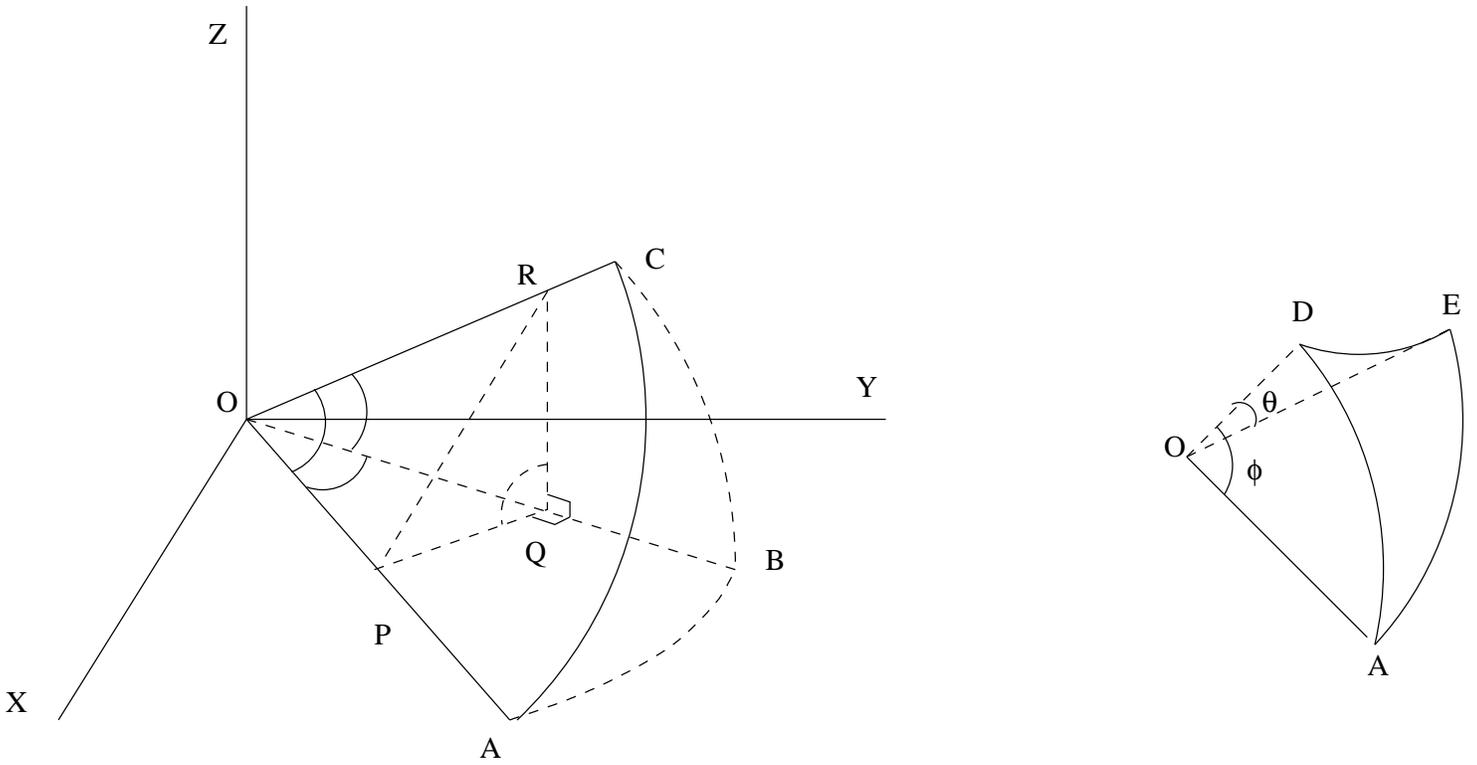}}
\caption{Two successive finite rotations}
\end{figure}

\begin{figure}
\centerline{
\epsfbox{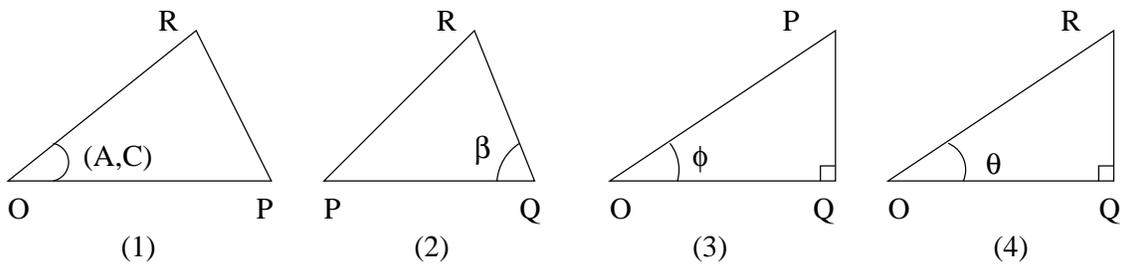}}
\caption{The Four Triangles}
\end{figure}

\end{document}